\documentclass[12pt]{article}
\usepackage[dvips]{graphicx}
\usepackage{graphicx}
\begin{document}
\title{Anomalous Magnetic Moment of W-boson at high temperature}
\author{ A.V.Strelchenko\\
Department of Physics, Dniepropetrovsk National University\\
St. Naukova 13, Dniepropetrovsk  49050, Ukraine\\
strel@ff.dsu.dp.ua}
\date{}
\maketitle
\thispagestyle{empty}
\begin{abstract}
By the Schwinger proper-time method, the one-loop contribution to the
anomalous magnetic moment of the W-boson is calculated at high
temperature. It is shown that the value of AMM is positive and
depends linearly upon temperature.

\end{abstract}

\section{Introduction}

~

 The investigation of high-temperature radiative corrections to
various physical values under the influence of external fields
remains one of the important topics in high energy physics and
cosmology. In particular, this makes it possible to explain some
mechanism of the spontaneous generation of primordial magnetic field
in the early Universe. Thus, in the paper \cite{Scl} it  was found that
the radiative correction to the W-boson ground state, for
$T\gg\sqrt{e H}$, stabilizes the vacuum of gauge bosons. Hence,
one comes to self-consistent picture when, at high temperature,
the spontaneously arisen  magnetic field is stabilized by
the radiative mass of charged gauge fields.

 In the present paper we calculated a one-loop correction to the
  anomalous magnetic moment of the W-boson at high temperature. For
  this purpose the Schwinger proper-time method was applied  which is quite
   appropriate  at high-temperature region,  where it is sufficient to take
    into account a static limit \cite{Kal}.

\section{Mass Operator}

~

Let us consider the Lagrangian of the electroweak theory
(omitting  fermions):
\begin{equation}\label{one}
L_{WS}=L_{Gauge} + L_{FP}+L_{Scalar},
\end{equation}
where

\newpage
\begin{eqnarray}
&L_{Gauge}=-\frac{1}{4}{F_{\mu\nu}}^2 -\frac{1}{4}{Z_{\mu\nu}}^2
-W^{+\mu}((\widetilde{P}^{2}-M^{2})g_{\mu\nu}-\widetilde{P}_{\mu} \widetilde{P}_{\nu}
-2 \imath e (\widetilde{F}_{\mu\nu}+F_{\mu\nu}^{R}))W^{-\nu} &\nonumber\\
& +\frac{1}{2}M_{Z}^{2}-g^{2}\cos^{2}{\theta_{W}}(Z_{\mu}Z^{\mu}W_{\nu}^{+}W^{-\nu}
-W_{\mu}^{+}Z^{\mu}W_{\nu}^{-}Z^{\nu})~~~~~~~~~~~&\nonumber\\
&~~~~~~+\frac{1}{2}\imath g \cos{\theta_{W}}((P_{\mu}W_{\nu}^{+}-P_{\nu}W_{\mu}^{+}))
(W^{-\mu}Z^{\nu}-W^{-\nu}Z^{\mu})-conjugated &\nonumber\\
&~~~+\imath g \cos{\theta_{W}}{Z_{\mu\nu}}W^{+\mu}W^{-\nu}
-\frac{g^{2}}{2}((W_{\mu}^{+} W^{+\mu})^{2}-W_{\mu}^{+} W^{+\mu}
W_{\nu}^{-} W^{-\nu} ),&\nonumber
\end{eqnarray}
\begin{eqnarray}
&L_{FP}=\overline{C}_{W}(P^{2}-M^{2})C_{W}- e (Z_{\mu}\cot{\theta_{W}}-A_{\mu})
\overline{C}_{W}P_{\mu}C_{W}~~~~~~~~~~~~~~~~~~~~~~~~~~~~~~~~~ &\nonumber\\
&+\imath e W_{\mu}^{-}\overline{C}_{W}\partial^{\mu}
(C_{Z}\cot{\theta_{W}}- C_{A})
-\frac{M}{2}g (\phi+\imath \varphi_{Z}) \overline{C}_{W}C_{W}+~~~~~~~~~~~~~~ &\nonumber\\
&+ \frac{\imath g}{2}M_{Z}\varphi^{-}\overline{C}_{Z}C_{Z}+ conjugated -
\overline{C}_{A}\partial^{2}C_{A}-~~~~~~~~~~~~~~~~~~~~~~~~~~~~~~&\nonumber\\
&-\overline{C}_{Z}(\partial_{\mu}^{2}+M_{Z}^{2})C_{Z}
-\frac{g M_{Z}\phi}{2\cos{\theta_{W}}}\overline{C}_{Z}C_{Z}
-\frac{\imath g}{2}M_{Z}\overline{C}_{Z}(C^{\ast}_{W}\varphi^{-}-C_{W}\varphi^{+})~~~~~~&\nonumber\\
&-\imath e (\partial_{\mu}\overline{C}_{Z}\cot{\theta_{W}}-
\partial_{\mu}C_{A})(W_{\mu}^{+}C_{W}-W^{-\mu}C^{\ast}_{W}),~~~~~~~~~~~~~~~~~~~~~~~~~~~ &\nonumber
\end{eqnarray}
and the exact expression of the scalar lagrangian $L_{Scalar}$ can be found, for example,
in \cite{CL}.
Here $W^{\pm}_{\mu}$,$Z_{\mu}$ and $A_{\mu}$ are the usual vector gauge boson fields,
$M$ and $M_{Z}$ represent  the masses of the W-boson and Z-boson, respectively
,$\varphi^{\pm}$ and $\varphi_{Z}$ are goldstone bosons,$\phi$ is Higgs scalar,
$ C_{W}$,$ C^{\ast}_{W}$,$ C_{Z}$,$C_{A}$ are the Faddeev-Popov`s ghosts.
The external magnetic field is introduced by dividing the potential
$A_{\mu}$ into radiative and classical parts, $A_{\mu}= A_{\mu}^{R}+\widetilde{A}_{\mu}$.
We introduced the following standard designations:
 $ F_{\mu\nu}=\partial_\mu A_\nu-\partial_\nu A_\mu  ,$
$  Z_{\mu\nu}=\partial_\mu Z_\nu-\partial_\nu Z_\mu ,$
$ {P}=\imath \partial_{\mu}+e {A}_{\mu} $ and $ \widetilde{P}=\imath \partial_{\mu}+
e \widetilde{A}_{\mu} $ are the covariant derivations,
 $ e=g\sin{\theta_{W}}$.
 It is convenient to choose the external potential in
the form $\widetilde{A}_{\mu}= (0,0,H x,0)$, $H=const$.

In the one-loop approximation, the W-boson mass operator is
determined by the standard set of diagrams in Fig.1 \cite{Van}, where
double lines represent the Green's function of charged particles:
solid double lines shaded inside are for the W-boson Green's
function, solid double lines painted out inside are for the Green's
function of the Goldstone particle $\varphi^{\pm}$, dashed
double lines shaded inside are for the Green's
function of the charged ghost components $C^{\pm}$. Thin wavy line corresponds to
 a radiative photon $A_{\mu}^{R}$ and Z-boson,
thin solid line corresponds to a neutral Higgs scalar $\phi$, and thin dashed line corresponds
to  a neutral ghost component $C_{A}$ and $C_{Z}$, respectively. The W-boson mass operator
in a magnetic field at high temperature can be written as (in the Feynman gauge)

\begin{equation}\label{two}
M_{\mu\nu}=\frac{e^2}{\beta}\int\frac{d^3k}{(2\pi)^3}
\left[M^\phi_{\mu\nu}(k,p)+M^{W}_{\mu\nu}(k,p)+M^{Z}_{\mu\nu}(k,p)\right],
\end{equation}
where

\[
M^\phi_{\mu\nu}(k,p)=\frac{1}{4\sin^{2}{\theta_{W}}}(k^2+m^2)^{-1}\left[(2k-p)_\mu
G(p-k)(2k-p)_\nu-4 M^2 G_{\mu\nu}(p-k)\right]
\]
\[
+\frac{1}{4\sin^{2}{\theta_{W}}}(k^2+M_{Z}^2)^{-1}\left[(2k-p)_\mu
G(p-k)(2k-p)_\nu\right],
\]
\begin{eqnarray*}
M^W_{\mu\nu}(k,p)&=&k^{-2}\Bigl\{\Gamma_{\mu\alpha,\rho}
G_{\alpha\beta}(p-k)\Gamma_{\nu\beta,\rho}+(p-k)_\mu
\Delta(p-k)k_\nu+ \\
& &+k_\mu\Delta(p-k)(p-k)_\nu+M^2 \delta_{\mu\nu}G(p-k)+ \\
& &+\frac{1}{4\sin^{2}{\theta_{W}}}k^2 \Bigl[G_{\mu\nu}(p-k)-2G_{\mu\nu}(p-k)+
\delta_{\mu\nu}(G_{\rho\rho}(p-k)+1)\Bigr]\Bigr\},\\
M^Z_{\mu\nu}(k,p)&=&(k^2+M_{Z}^2)^{-1}\Bigl\{\cot^{2}{\theta_{w}}\Bigl[\Gamma_{\mu\alpha,\rho}
G_{\alpha\beta}(p-k)\Gamma_{\nu\beta,\rho}+(p-k)_\mu
\Delta(p-k)k_\nu+ \\
& &+k_\mu\Delta(p-k)(p-k)_\nu\Bigr]+\tan^{2}{\theta_{w}}M^2 \delta_{\mu\nu}G(p-k)\Bigr\},\\
\Gamma_{\mu\alpha,\rho}&=&\delta_{\mu\alpha}(2p-k)_\rho+\delta_{\alpha\rho}
(2k-p)_\mu+\delta_{\mu\rho}(p+k)_\alpha,
\end{eqnarray*}

$$ G_{\mu\nu}(P)=-[P^2+M^2+2ieF_{\mu\nu}]^{-1} $$ is the W-boson Green's
function,
$$G(P)=\Delta(P)=-[P^2+M^2]^{-1}$$ is the Green's
function of the Goldostone particlaes $\varphi^{\pm}$ and of the charged
ghost components, respectively,
$\beta=\frac{1}{T}$. For static (high temperature) limit we put
$k_{4}=0$ \cite{Kal}.

In order to evaluate this expression we used the Schwinger proper
time method, modified for the case of the high temperature
(see for details \cite{Scl}). Thus, the mass operator averaged over physical
states of a vector particle can be obtained in the form
\begin{eqnarray}\label{three}
&\langle n,\sigma \mid M_{ij}^\phi \mid
n,\sigma\rangle =\frac{2\alpha}{4\sin^{2}{\theta_{W}}\sqrt{\pi}\beta}
\int\limits_0^1 \frac{d u}{\sqrt{u}}
\int\limits_0^\infty \frac{d x}{\sqrt{x}}
[eH\Delta]^{-1/2}\exp\left[-xu\frac{M^2}{eH}\right]~~~~~~~~~~~~~~~~~~~~~&\nonumber\\
&~~~~~~\times\exp\Bigl\{-(2n+1)[\rho-x(1-u)]-2y(1-u)\Bigr\}&  \nonumber \\
&~~~~~~\times
(\exp\left[-\frac{x(1-u)}{u}
\frac{m^2}{eH}\right](K(y)-M^2e^{2y})+
\exp\left[-\frac{x(1-u)}{u}\frac{M^2_{Z}}{eH}\right]K(y)),&
\end{eqnarray}
\begin{eqnarray}\label{four}
&\displaystyle\langle  n,\sigma \mid M_{ij}^W \mid
n,\sigma\rangle =\frac{\alpha}{2\sqrt{\pi}\beta} \int_0^1 \frac{d u}{\sqrt{u}} \int_0^\infty
\frac{d x}{\sqrt{x}}(eH\Delta)^{-1/2}\exp\left[-xu\frac{M^2}{eH}\right]~~~~~~~~~~~~~~~~~~
& \nonumber \\
&~~~~~~~~~~~~~~~~\times \exp\Bigl\{-(2n+1)[(\rho-x(1-u)]-2y(1-u)\Bigr\}M(x,u),&
\end{eqnarray}
\begin{eqnarray}\label{five}
&\displaystyle\langle  n,\sigma \mid M_{ij}^Z \mid
n,\sigma\rangle =\frac{\alpha}{2\sqrt{\pi}\beta} \int_0^1 \frac{d u}{\sqrt{u}} \int_0^\infty
\frac{d x}{\sqrt{x}}(eH\Delta)^{-1/2}
\exp\left[-xu\frac{M^2}{eH}-\frac{x(1-u)}{u}\frac{M^2_{Z}}{eH}\right]  &\nonumber \\
&\times \exp\Bigl\{-(2n+1)[(\rho-x(1-u)]-2y(1-u)\Bigr\}&\nonumber\\
&~~~~~~~\times (\cot^{2}{\theta_{W}}\widetilde{M}(x,u)+
\tan^{2}{\theta_{W}}M^{2}+2\cot^{2}{\theta_{W}}K(y)),&
\end{eqnarray}
where $x=euHs$, $y=x\sigma$,

$$\tanh \rho=\frac{(1-u)\sinh x}{(1-u)\cosh x+u\frac{\sinh x}{x}},$$
$$\Delta=(1-u)^2+2u(1-u)\frac{\sinh 2x}{2x}+u^2\frac{\sinh ^2x}{x^2}.$$
The explicit expressions for
$M(x,u)$, $\widetilde{M}(x,u)$ and $K(y)$ are given in Appendix.

\section{Anomalous Magnetic Moment}
Now consider the case of  weak magnetic fields which are characterized
by the condition $\frac{e H}{M^{2}} \ll 1$. As is known, the one-loop
contribution to the anomalous magnetic moment of the W-boson is
defined as \cite{Van}
\begin{equation}\label{six}
\Delta \kappa = -\frac{1}{2} \frac{\partial Re<M>}{\partial e H
\sigma} \mid_{H=0}.
\end{equation}
To perform the integrations in the Eqs. (\ref{three}) and (\ref{four}) it is
necessary to divide the integral region over $u$ into the two parts:
$0\leq u < u_{0}$ and $ u_{0} < u \leq 1$, where
$\frac{e H}{M^{2}} \ll  u_{0} \ll 1$. In the region $ u_{0} < u \leq 1$
only small values of $x$ contribute, and one can expand the
integrands on the r.h.s. into the power
series of $x$. In a similar manner, in the region $0\leq u < u_{0}$,
one can use the expansion into the power series of $u$. The result
should not depend on $u_{0}$. After carrying out the integrations over
 $u$ and $x$  we obtain for scalar sector (diagrams a.- b.)
\begin{equation}\label{seven}
\Delta \kappa_{\phi} = 0.212 \frac{T}{M} \alpha,
\end{equation}
and for gauge sector (diagrams c.- h.)
\begin{equation}\label{eight}
\Delta \kappa_{W} = 1.048 \frac{T}{M} \alpha,
\end{equation}
\begin{equation}\label{nine}
\Delta \kappa_{Z} = 2.121 \frac{T}{M} \alpha.
\end{equation}
Eqs. (\ref{seven}) , (\ref{eight}) and (\ref{nine}) yield the bosonic contribution to the
W-boson AMM in the high-temperature approximation.

\newpage

\section{Conclusion}
~
The anomalous magnetic moment of the W-boson at zero temperature was
calculated in \cite{Van} and can be written in the following way:
\begin{equation}\label{ten}
\Delta \kappa_{W} = \frac{7}{16} \frac{e^{2}}{\pi^{2}} ,
\end{equation}
\begin{eqnarray}\label{eleven}
\Delta \kappa_{\phi} = \frac{g^{2}}{16 \pi^{2}}
\int\limits_0^1 \frac{d u u^{2}}{u^{2}+\frac{m^{2}}{M^{2}}(1-u)}(u^{2}-u+
2+\frac{m^{2}}{2M^{2}}(1-u)) ,
\end{eqnarray}
\begin{eqnarray}\label{twelve}
\Delta \kappa_{Z} &=& \frac{g^{2}\cos^{4}{\theta_{W}}}{16 \pi^{2}}
\int\limits_0^1 \frac{d u u^{2}}{u^{2}\cos^{2}{\theta_{W}}+(1-u)}(8u-10u^{2}+12u^{3}-\nonumber\\
&-&\frac{1}{\cos^{2}{\theta_{W}}}(2-4u+9u^{2}-u^{3})-\frac{1}{2\cos^{4}{\theta_{W}}}(4-5u-u^{2})).
\end{eqnarray}
 The result obtained in the present paper  represents a
high-temperature correction to these quantities.
As is directly followed from Eqs. (\ref{seven}) , (\ref{eight}) and (\ref{nine}), at high
temperature and sufficiently weak magnetic fields,
when $\sqrt{eH} \ll T \ll M$, the value of the anomalous
magnetic moment is positive and increases linearly with
the growth of temperature.

It is interesting to compare this result with the AMM of the
fermion, calculated in \cite{Ter}(see also review article  \cite{Bor}).
Therein it is found that the temperature correction to the
gyromagnetic ratio of the electron has the form
$a_{T}=\frac{\alpha\pi}{9}(\frac{T}{m_{e}})^{2}$.
As one can see, the temperature contribution to the fermion AMM
depends upon $T^{2}$ term, whereas in the
W-boson case the AMM is proportional to $T$. In contrast to the
fermion case, where contribution to the magnetic moment
can be separated at any values of the field $H$,
the radiative correction to the W-boson AMM can be
defined only in weak fields.

\section{Acknowledgements}
The author thanks to V.V. Skalozub for proposing the problem, helpful
discussions  and reading the manuscript.

\newpage
\section{Appendix}

The explicit expressions obtained upon averaging the mass operator of the W-boson
over physical states have the forms
\begin{eqnarray}
&M(x,u)=\widetilde{M}(x,u)-2 K(y)+\frac{1}{\sin^{2}{\theta_{\theta_{W}}}}
(\cosh ^2{x}-2e^{-2y})\Biggl\{up^2_3+(2n+1)eH\frac{u^2\sinh ^2{x}}{x^2\Delta}~~~~~~~~~&\nonumber \\
&+eH\frac{u^2}{4x^2\Delta}\Bigl[2\sinh{2x}+\frac{4x(1-u)}{u}\Bigr]+
\frac{ueH}{2x}\Biggr\},~~~~~~~~~~~~~~~~~~~~~~~~~~~~~~~~~~~~~&\nonumber
\end{eqnarray}
\begin{eqnarray}
&\widetilde{M}(x,u)=2e^{2y}\Biggl\{neH\left[8\sinh ^2{x}-\frac{u^2}{x^2\Delta}\left(
\frac{4x(1-u)}{u}\cosh {3x}\sinh{x}+8\sinh {2x}\sinh ^2{x}\right)\right]~~~~~~~~~~~~~~~~~~~~~~&\nonumber \\
&+eH\left[2A(x)-\frac{u^2A(x)}{2x^2\Delta} \left(A(x)-
\frac{x(1-u)}{u}\left(A(x)+2\cosh {2x}\right)\right)\right]-
2up^2_3+2\bar{p}^2\Biggr\}~~~~~~~~~~&\nonumber \\
&+2e^{2y}\Biggl\{up^2_3+(2n+1)eH
\frac{u^2\sinh ^2{x}}{x^2\Delta}
+eH\frac{u^2}{4x^2\Delta}\Bigl[2\sinh{2x}+\frac{4x(1-u)}{u}\Bigr]+
\frac{ueH}{2x}\Biggr\}~~~~~~~~~~~&\nonumber \\
&+eH\Biggl\{2A(-y)\frac{u(1-u)}{x}\frac{\cosh{2x}}{\Delta}
+\frac{u^2(8\sinh ^2{x}+1)}{2x^2\Delta}\left[\sinh {2x}+\frac{4x(1-u)}{u}\right]
+\frac{u}{2x}\Biggr\}~~~~~~~~~~~~~~~~~~&\nonumber \\
&+neH\Biggl\{4\sinh ^2{x}+N(x)+N(-x)+\frac{u^2\sinh ^2{x}}{x^2\Delta}\left[2\cosh {2x}+
\frac{4x(1-u)}{u}\sinh {2x}\right]\Biggr\}~~~~~~~~&\nonumber \\
&+(2n+1-\sigma)eH
\Biggl\{\frac{A^2(y)+(1-4\sinh ^2{y})e^{-2y}D(-y)+2\frac{y}{u}
(1-u)A(y)}{D(y)}\left[\frac{A(y)}{D(y)}-u\right]~~~~~~~~~~~~~~~~~~~&\nonumber \\
&+\frac{u^2\sinh ^2{y}}{y^2\Delta(y)}
\left[A(y)+ue^{-2y}D(y)\right]
[1-D(-y)]+\frac{u}{2}
\Biggr[ 1+e^{2y} +\frac{ u^2 \sinh ^2{y}}{y^2 \Delta(y)}D(-y)\Biggl]~~~~~~~~~~~&\nonumber \\
&+\left[\frac{4(1-u)u\sinh ^2{y}}{y\Delta(y)}-\frac{u^2\sinh ^2{y}}{y^2\Delta(y)}+
u\frac{A(y)}{D(y)}\right]
\left[A(-y)+e^{2y}y\frac{1-u}{u}\right]~~~~~~~~~~~~~~~~~~~~~~~~~~~~~~~~~~&\nonumber \\
&-\left[(1+e^{2y})
\frac{A(-y)}{D(-y)}+A(-y)\frac{u^2\sinh ^2{y}}{y^2\Delta(y)}\right]
\left[\frac{1}{2}-\frac{y}{u}(1-u)\right]\Biggr\}+u(2+u)p^2_3+\overline{p}^{2}~~~~~~~~~~~~~~&\nonumber\\
&+(10+8 \sinh ^2{y})K(y)+\frac{eH\sigma}{D(-y)}\Bigl[ 8A(y)-2A(-y)-
A^2(-y)-2A^2(y)\Bigr]~~~~~~~~~~~~~~~&\nonumber\\
&+eH \Biggl\{ A(x)-\frac{A(x)}{D(-x)}+e^{-2x} \Bigl[1-
\frac{2x}{u}(1-u)\Bigr] \Bigl[ \frac{A(x)}{D(x)}+\frac{u^2\sinh ^2{x}}{x^2 \Delta}
\Bigr] \Biggr\},~~~~~~~~~~~~~~~~~~~~~~~~~~&\nonumber
\end{eqnarray}
where
\begin{eqnarray}
&\Delta(y)=(1-u)^2+2u(1-u)\frac{\sinh {2y}}{2y}+u^2\frac{\sinh ^2{y}}{y^2},
\qquad \Delta \equiv \Delta(x),~~~~~~~~~~~~~~~~~~~~~~~& \nonumber \\
&K(y)=\frac{1}{2}(2n+1-\sigma) \Bigl[-u^2 e^{-2y} \frac{D(-y)}{D(y)}-2u \frac{A(y)}{D(y)}+
\frac{u^2 \sinh ^2{y}}{y^2 \Delta(y)} \Bigr] +\frac{eH \sigma}{\Delta(-y)},~~~~~~~~~~~~~~~~&\nonumber\\
&N(x)=\frac{A(x)}{D(x)} \Bigl[ (2\cosh {2x}-\frac{e^{-2x}2x}{u}(1-u)\Bigr],
~~~~~~~~~~~~~~~~~~~~~~~~~~~~~~~~~~~~~~~~~~~~~~~~~ &\nonumber\\
&A(y)=e^{-2y}-1, \qquad D(y)=A(y)-\frac{2y}{u}(1-u).
~~~~~~~~~~~~~~~~~~~~~~~~~~~~~~~~~~~~~~~~&\nonumber
\end{eqnarray}
\newpage
\begin{figure}\centering
\includegraphics[scale=0.7]{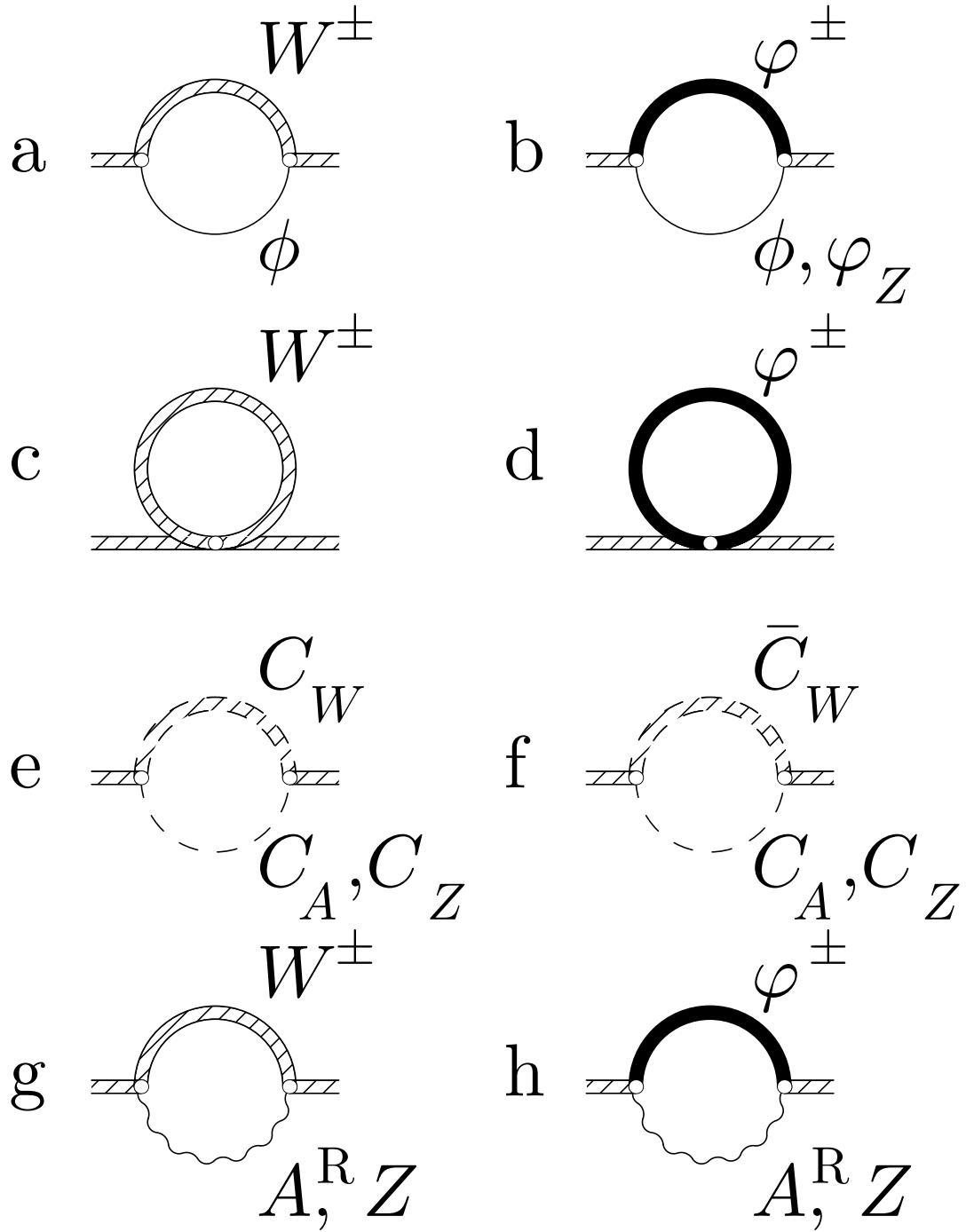}
\caption{W-boson mass operator in the one-loop approximation.} \label{Fig1}
\end{figure}

\end{document}